\documentclass[preprint,12pt    ]{elsarticle}
\usepackage{graphicx}
\usepackage{float}
\usepackage{latexsym,amsbsy,epsfig,fancybox}
\usepackage{amstext}
\usepackage{subfigure}
\usepackage{amsfonts,textgreek}
\usepackage{mathrsfs}
\usepackage{algpseudocode}
\usepackage[most]{tcolorbox}
\usepackage{mathtools}
\usepackage{tikz}
\usepackage{pstricks}
\usepackage{array}
\usepackage{amssymb}
\usepackage{multirow}
\usepackage{booktabs}
\usepackage{verbatim}
\usepackage{natbib}
\biboptions{sort&compress}
\usepackage{url}
\usepackage{xr}
\usepackage[margin=3cm]{geometry}
\usepackage{soul}
\usepackage[colorlinks=true,citecolor=blue]{hyperref}

\newcommand{\ac}[1]{\textcolor{red}{add citation}}

\newcommand{\af}[1]{\textcolor{red}{add figure}}
\newcommand{\ar}[1]{\textcolor{red}{add crossref}}
\tikzstyle{nicebox}=[draw=black!100, fill=white!10, rectangle, inner sep=4pt, inner ysep=16pt]
\tikzstyle{niceboxtitle}=[draw=black!100, fill=white, text=black, rectangle]

\usetikzlibrary{arrows,decorations.markings}

\journal{arXiv}
\begin{document}
\begin{frontmatter}

\title{From Connectivity to Rupture: A Coarse-Grained Stochastic Network Dynamics Approach to Polymer Network Mechanics}
\author[1]{Shaswat Mohanty}
\author[1]{Wei Cai\corref{cor1}}
\ead{caiwei@stanford.edu}
	
\address[1]{Department of Mechanical Engineering, Stanford University, CA 94305-4040, USA}

\cortext[cor1]{Corresponding author}

\begin{abstract}
We introduce a coarse-grained stochastic network dynamics (CGSND) framework for modeling deformation and rupture in polymer networks. The method replaces explicit molecular dynamics (MD) or coarse-grained molecular dynamics (CGMD) with network-level evolution rules while retaining chain entropic elasticity and force-controlled bond failure. Under uniaxial loading, CGSND reproduces the characteristic nonlinear stress--stretch response of elastomeric networks, including a well-defined ultimate tensile strength and post-peak softening due to progressive bond rupture. Comparison with coarse-grained molecular dynamics (CGMD) simulations shows that CGSND captures the qualitative form of the stress response and the onset of catastrophic damage despite its rate-independent formulation. Analysis of rupture kinetics reveals a pronounced peak in the bond-breaking hazard rate near the ultimate tensile strength in both approaches. In addition, the distribution of broken segment lengths remains statistically indistinguishable from the initial network, indicating that rupture is not biased toward short or long chains. Finally, the evolution of the Gini coefficient of bond force magnitudes reveals strong force localization preceding failure. These results demonstrate that CGSND provides a computationally efficient and physically interpretable framework for connecting force localization and rupture kinetics to macroscopic failure in polymer networks.
\end{abstract}

\begin{keyword}
Polymer Network \sep Rupture Strength \sep coarse-grained stochastic network dynamics \sep Selective Bond-breaking
\end{keyword}

\end{frontmatter}

\section{Introduction}

Cross-linked polymer networks form the mechanical backbone of a wide class of elastomers, and hydrogels~\cite{aprem2005recent,maitra2014cross}. Their macroscopic response arises from the collective deformation, load redistribution, and rupture of a disordered network of polymer strands. Despite decades of study, connecting microscopic chain-level mechanics to macroscopic failure remains a central challenge in polymer physics \cite{clarke1987fracture,caceres1995deformation,berry1961fracture,kendall1983relation}.

Coarse-grained molecular dynamics (CGMD) simulations provide a powerful route to studying polymer network deformation and fracture by explicitly resolving chain elasticity and bond rupture \cite{mohanty2025strength, yin2024network}. However, their high computational cost limits accessible system sizes and statistical sampling, and the complexity of the resulting dynamics can obscure higher-level organizing principles governing load transfer and failure \cite{joshi2021review}.

An alternative perspective is to represent polymer networks as weighted graphs, with the monomeric units as nodes and the covalent bonds (Kuhn segments) as edges that carry mechanical load \cite{yin2020topological, yin2024network, mohanty2025strength}. Within this framework, deformation corresponds to the evolution of the connectivity network, while fracture emerges from progressive edge removal. More generally, studies of evolving networks have demonstrated that coarse-grained stochastic descriptions—based on aggregate observables rather than explicit microscopic dynamics—can capture essential system behavior at greatly reduced computational cost~\cite{katsoulakis2003coarse,seiferth2020coarse}. Whether such approaches can reliably reproduce mechanically driven failure processes, where force localization and extreme events dominate, remains an open question.

In this work, we introduce a coarse-grained stochastic network dynamics (CGSND) framework for modeling deformation and rupture in polymer networks. The method replaces explicit molecular dynamics with network-level evolution rules while retaining nonlinear entropic elasticity and force-controlled bond failure. We apply CGSND to uniaxial deformation of cross-linked polymer networks and compare its predictions directly with CGMD simulations. We show that CGSND reproduces the qualitative stress--stretch response and the onset of catastrophic damage, captures the kinetics of bond rupture through a well-defined hazard-rate peak, and reveals strong force localization preceding failure via the Gini coefficient of bond force magnitudes.

The remainder of the paper is organized as follows. Section~\ref{sec:Methodsp3} describes the coarse-graining strategy and stochastic network dynamics formulation. Section~\ref{sec:Resultsp3} presents numerical results, including comparisons with CGMD and analysis of stress response, rupture kinetics, and force localization. Section~\ref{sec:Discp3} discusses the implications and limitations of the approach, and Section~\ref{sec:Concp3} concludes with perspectives on extensions and applications.

\section{Methodology} \label{sec:Methodsp3}

\subsection{Coarse-Graining of the Polymer Network}

The starting point of the model is a coarse-grained molecular dynamics (CGMD) representation of a cross-linked polymer network. To construct the graph-based representation, only the cross-linked beads are retained as graph nodes. An undirected edge is created between two nodes if the corresponding beads are connected either by a chemical cross-link or by a polymer strand (sequence of monomer units along the polymer backbone). Each edge is assigned an integer weight $w_{ij}$ corresponding to the number of bonds in the connecting strand, or unity if the connection is a direct cross-link~\cite{yin2020topological, yin2024network, mohanty2025strength}.

In addition to the edge weights, two other quantities are tracked for each edge during the deformation process: (i) the inter-node distance, computed from the coordinates of the two beads subjected to periodic boundary conditions, and (ii) the normalized extension, defined relative to the edge contour-length scale. These quantities, together with the evolving network topology, are sufficient for evaluating strand forces and rupture events. Unlike prior graph-based approaches~\cite{koplik1981effective,sax1983modeling,brosseau2002generalized}, no path-based or effective-medium reductions are performed; instead, the network evolves directly under applied deformation by tracking its explicit connectivity.

\subsection{Loading and deformation}

Uniaxial deformation is imposed kinematically by applying an affine stretch to node coordinates along the loading axis. Lateral coordinates evolve consistently with the imposed boundary conditions (volume-preserving stretch in the present work). At each strain increment, bond end-to-end vectors $\mathbf D_{ij}$ are updated using periodic boundary conditions.

\paragraph{Normalized extension}
Each edge $(i,j)$ carries an integer weight $w_{ij}$ equal to the number of bonds in the connecting strand (unity for direct cross-links). We associate a contour-length scale
\[
L_{ij} = \alpha\, w_{ij},
\]
where $\alpha\sim1.3$ is a constant mapping factor from bond count to contour length, calibrated from the maximum allowable enthalpic stretch from CGMD model~\cite{mohanty2024understanding,zhang2024modeling}. The normalized extension is
\[
p_{ij} = \frac{\|\mathbf D_{ij}\|}{L_{ij}} .
\]

\paragraph{Bond force law}
Bond forces are modeled using an inverse-Langevin stiffened spring law~\cite{jedynak2017new}. The inverse Langevin function is approximated by
\begin{equation}
\mathcal{L}^{-1}(p) \;=\; \frac{p \left( 3 - 1.00651 p^2 - 0.962251 p^4 + 1.47353 p^6 - 0.48953 p^8 \right)}{(1 - p)(1 + 1.01524 p)} ,
\end{equation}
and the nondimensional bond force vector is defined as
\begin{equation}
\mathbf{f}_{ij}^{*} = \mathcal{L}^{-1}(p_{ij}) \, \frac{\mathbf{D}_{ij}}{\|\mathbf D_{ij}\|} .
\end{equation}
This choice recovers a linear spring response at small extension and exhibits strong strain stiffening as $p_{ij}\to 1$.
We note that $\mathcal{L}^{-1}(p_{ij})$ acts here as a stretch-dependent stiffness multiplier; the form is intended as a computationally efficient entropic stiffening law rather than an exact freely-jointed chain relation.

\paragraph{Bond rupture}
Bonds are removed irreversibly when $\|\mathbf{f}_{ij}^{*}\| > f_{\mathrm{cut}}^{*}$, producing progressive degradation of the network backbone. The rupture force for a bond is set at $f_{\mathrm{cut}}^{*}\sim1431.65$ which corresponds to a rupture stress of $10$ GPa~\cite{yang2019polyacrylamide, helaly2011effect, wang2023research}.

\paragraph{Stress calculation (reported in MPa)}
Macroscopic stress is computed via a bulk virial formulation~\cite{tsai1979virial},
\begin{equation}
\boldsymbol{\sigma}^{*} = \frac{1}{V^{*}} \sum_{(i,j)} \mathbf{D}_{ij} \otimes \mathbf{f}_{ij}^{*} ,
\end{equation}
where $V^{*}$ is the instantaneous simulation volume in reduced length units, $\mathbf{D}_{ij}$ is the bond end-to-end vector, and $\mathbf{f}_{ij}^{*}$ is the corresponding nondimensional bond force.

To report stress in physical units, we introduce a coarse-graining length scale $b$ corresponding to the Kuhn length of the polymer and define a reference volume $V_0 = b^3$. The physical stress is then obtained as
\begin{equation}
\boldsymbol{\sigma}\;[\mathrm{MPa}] = 10^{-6}\,\frac{k_B T}{V_0}\,\boldsymbol{\sigma}^{*},
\end{equation}
where $k_B$ is Boltzmann’s constant and $T$ is temperature. In the present work we use $b = 8.6\times 10^{-10}\,\mathrm{m}$~\cite{habenschuss2007structure}, such that $V_0 = b^3$, and we denote the component the stress in the loading direct by $S=\sigma_{xx}$.

\paragraph{Bond rupture hazard rate}
To characterize the rate at which damage accumulates during deformation, we compute the bond rupture hazard rate as a function of stretch~\cite{block1998reversed}. Let $N_0$ denote the total number of bonds in the undeformed network, obtained as the sum of edge weights
$N_0 = \sum_{(i,j)} w_{ij}$. At a given stretch $\lambda$, let $N_b(\lambda)$ denote the cumulative number of bonds that have ruptured.

We define the bond rupture hazard rate with respect to stretch as
\begin{equation}
h(\lambda) = \frac{1}{N_0}\,\frac{\mathrm d N_b}{\mathrm d \lambda},
\end{equation}
which represents the instantaneous probability per unit stretch that a bond fails, normalized by the initial bond population. In practice, $h(\lambda)$ is evaluated numerically using finite differences between successive strain increments.

Bond rupture events are further partitioned according to bond type. Each ruptured edge contributes a single bond failure, regardless of its weight, and is classified as either a cross-link bond ($w_{ij}=1$) or a backbone bond ($w_{ij}>1$). Separate hazard rates are therefore computed for cross-link bonds and backbone bonds, allowing direct comparison of rupture kinetics between different bond populations.

The hazard rate provides a strain-resolved measure of damage accumulation that is independent of absolute bond counts and remains well-defined even after substantial network degradation.

\paragraph{Force localization and Gini coefficient}
To quantify load-path concentration during deformation, we compute the Gini coefficient of bond force magnitudes~\cite{gini2016qsar,xia2019globule}. For a given deformation state, let $\{f_k\}_{k=1}^{N_b}$ denote the set of nondimensional force magnitudes $f_k = \|\mathbf f_{ij}^{*}\|$ over all intact bonds in the network, where $N_b$ is the number of surviving bonds. The Gini coefficient is defined as
\begin{equation}
G = \frac{1}{2 N_b^2 \bar f} \sum_{k=1}^{N_b} \sum_{\ell=1}^{N_b} \left| f_k - f_\ell \right|,
\end{equation}
where $\bar f = \frac{1}{N_b}\sum_{k=1}^{N_b} f_k$ is the mean bond force magnitude.

The Gini coefficient provides a scalar measure of force heterogeneity within the network. $G=0$ corresponds to perfectly homogeneous load sharing, while $G \to 1$ indicates extreme force localization in which a small subset of bonds carries the majority of the load.
In the present framework, the Gini coefficient is evaluated at each strain increment using the instantaneous distribution of bond force magnitudes and is used to characterize the emergence and evolution of dominant load-bearing paths during deformation and failure.

\subsection{Loading Algorithm}
\begin{tcolorbox}[title=CGSND Polymer Network Loading, colback=black!5!white, colframe=black!25!black, rounded corners]
\begin{algorithmic}[1]
\State \textbf{Input:} final stretch $\lambda_\text{end}$, increment $\Delta \lambda$, rupture threshold $f_{\mathrm{cut}}^{*}$
\State \textbf{Initialize:} set $\lambda \gets 1$; construct graph with edge weights $w_{ij}$; set initial node positions
\State \textbf{(Optional) Conditioning:}
    \For{$i = 1$ to $N_{\mathrm{cond}}$}
        \State update $\lambda \gets \lambda + \Delta\lambda$
        \State apply affine stretch to node positions
        \State update bond geometry and forces
        \State remove bonds with $\|\mathbf f_{ij}^{*}\| > f_{\mathrm{cut}}^{*}$
    \EndFor
\State \textbf{Loading to failure:}
    \For{$i = 1$ to $(\lambda_\text{end}-1)/\Delta \lambda$}
        \State update $\lambda \gets 1 + i\Delta\lambda$
        \State apply affine stretch to node positions
        \State update bond geometry and forces
        \If{limited relaxation enabled}
            \State perform $n_{\mathrm{relax}}$ force-driven relaxation steps (bond breaking disabled)
            \State recompute bond forces
        \EndIf
        \State remove bonds with $\|\mathbf f_{ij}^{*}\| > f_{\mathrm{cut}}^{*}$
        \State compute bulk virial stress:
        \[
        \boldsymbol{\sigma}^{*}=\frac{1}{V^{*}}\sum_{(i,j)} \mathbf D_{ij}\otimes \mathbf f_{ij}^{*}
        \]
        \State convert to MPa:
        \[
        \boldsymbol{\sigma}[\mathrm{MPa}] = 10^{-6}\,\frac{k_BT}{b^3}\,\boldsymbol{\sigma}^{*}
        \]
        \State record $S=\sigma_{xx}$, edge count, and weights (strand lengths) of broken edges
    \EndFor
\State \textbf{Output:} stress--stretch curve (MPa), bond rupture statistics, connectivity evolution
\end{algorithmic}
\end{tcolorbox}

\section{Results} \label{sec:Resultsp3}

We present the outcomes of the coarse-grained stochastic network dynamics (CGSND) framework applied to polymer network deformation under uniaxial loading and compare its predictions directly with coarse-grained molecular dynamics (CGMD) simulations of the same network. The results correspond to four complementary observables that together characterize the mechanical response, damage evolution, microscopic rupture statistics, and the force localization. For the stress--stretch response, bond rupture kinetics, and broken-segment length distributions, CGSND and CGMD results are shown side by side. Force localization is analyzed exclusively within CGSND, since extracting a comparable network-level measure from CGMD would require introducing arbitrary bond-averaging schemes that fall outside the scope of this study. Unless otherwise stated, stresses are reported in MPa using the thermal stress scale $k_B T / b^3$. The stress response is evaluated at strain increments of $\Delta \lambda = 0.01$, while network-level observables are computed at coarser increments of $\Delta \lambda = 0.1$ to reduce computational cost.

\subsection{Stress--stretch response}

\begin{figure}[!htbp]
    \centering
    \subfigure[]{\includegraphics[width=0.48\textwidth]{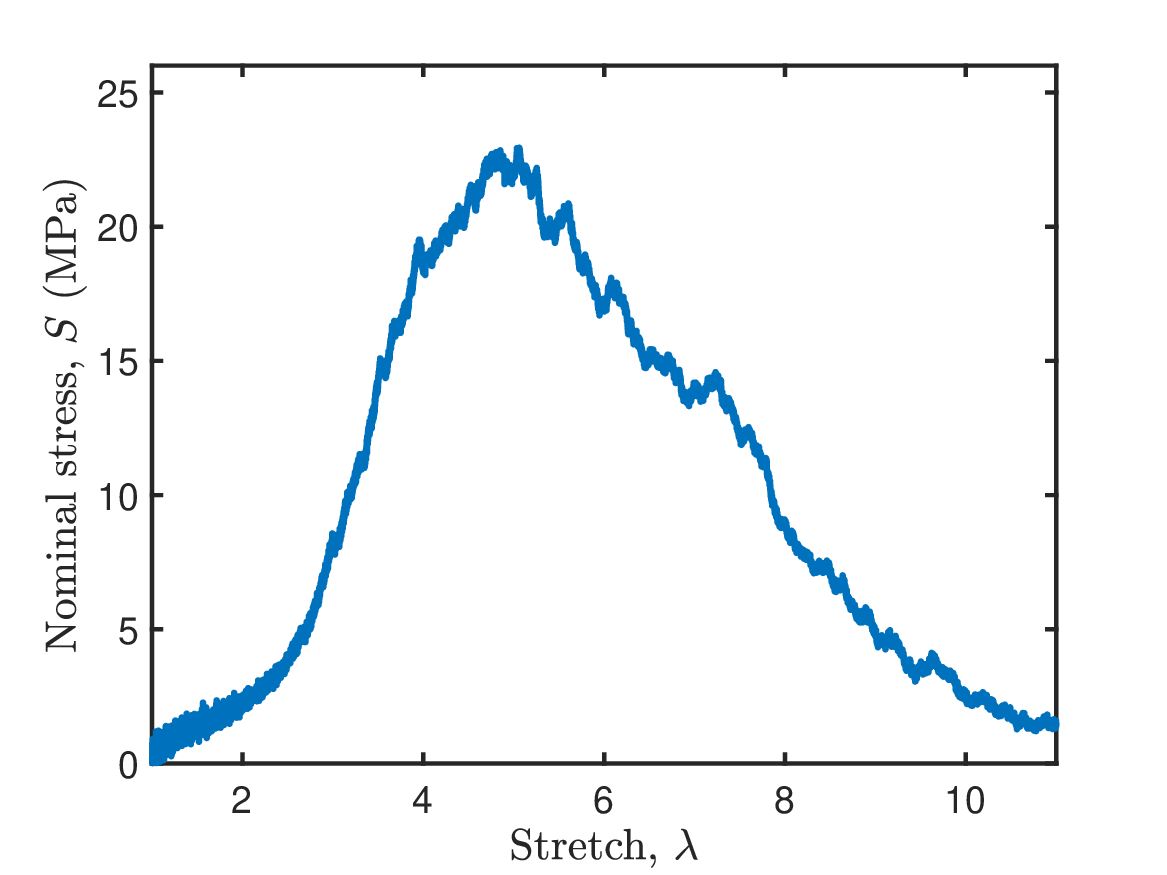}\label{fig:stress_cgmd}}
   \subfigure[]{\includegraphics[width=0.48\textwidth]{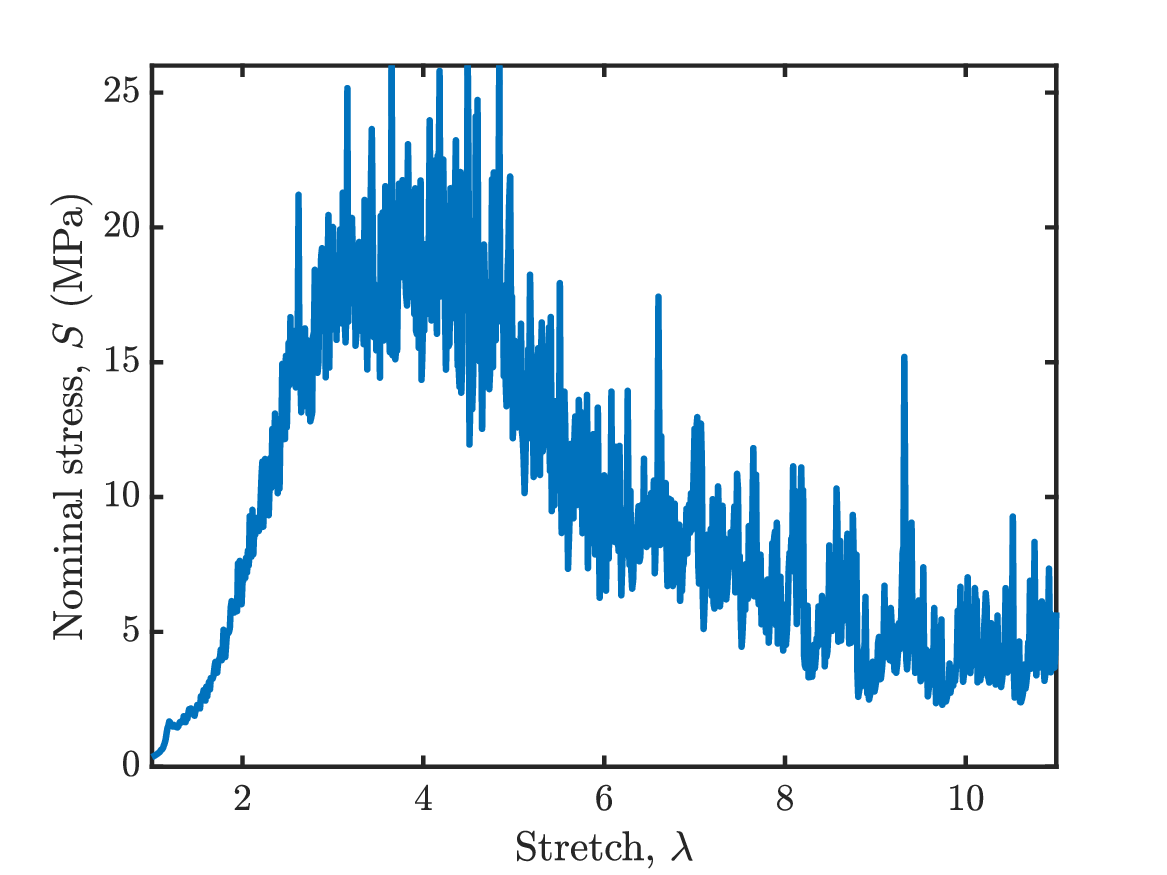}\label{fig:stress_cgsnd}}
    \caption{\textbf{Nominal stress--stretch response from CGMD and CGSND.}
Macroscopic nominal stress $\sigma$ as a function of applied stretch ratio $\lambda$ obtained from (a) CGMD simulations and (b) CGSND framework. Stress in CGSND is computed using a bulk virial formulation and reported in MPa via the thermal stress scale $k_B T / b^3$.}
    \label{fig:stress_stretch}
\end{figure}

Figure~\ref{fig:stress_stretch} compares the nominal stress--stretch response obtained from coarse-grained molecular dynamics (CGMD) simulations and the coarse-grained stochastic network dynamics (CGSND) framework. Both approaches exhibit the characteristic nonlinear mechanical response of elastomeric polymer networks. At small deformation, the stress increases gradually, reflecting entropic elasticity of network strands. With increasing stretch, the response transitions into a pronounced strain-stiffening regime associated with finite chain extensibility, and ultimately reaches a well-defined peak stress corresponding to the onset of macroscopic failure.

The CGSND framework reproduces the overall shape of the CGMD stress--stretch curve, including the location of the ultimate tensile strength and the subsequent post-peak softening regime. In both models, stress softening coincides with the accumulation of bond rupture events, indicating that mechanical failure is governed by progressive loss of load-bearing strands rather than a single catastrophic break.

Quantitative differences in stress magnitude and post-peak behavior are observed between the two approaches. While CGSND reproduces the overall shape of the CGMD stress–stretch response, including the location of the ultimate tensile strength, it exhibits a noisier signal and differences in peak stress magnitude and post-peak softening. These discrepancies arise from the affine, rate-independent nature of the CGSND formulation, which does not incorporate explicit dynamical relaxation, inertial effects, or thermostat-mediated dissipation present in CGMD simulations. As a result, CGSND represents a kinematically constrained limit of the molecular system, in which stress redistribution occurs primarily through discrete bond rupture and network reconfiguration rather than continuous nonaffine relaxation.

The stress response obtained from CGSND exhibits increased fluctuations relative to CGMD, especially in the post-peak regime. These fluctuations reflect the stochastic nature of bond breaking and the discrete rearrangement of load paths in the evolving network, whereas CGMD averages over thermal motion and relaxation processes on the timescale of deformation. Despite these differences, the close agreement in the qualitative stress--stretch behavior
demonstrates that the CGSND framework captures the dominant mechanisms governing nonlinear elasticity and fracture in polymer networks.

\subsection{Bond rupture kinetics}

\begin{figure}[!htbp]
    \centering
    \subfigure[]{\includegraphics[width=0.48\textwidth]{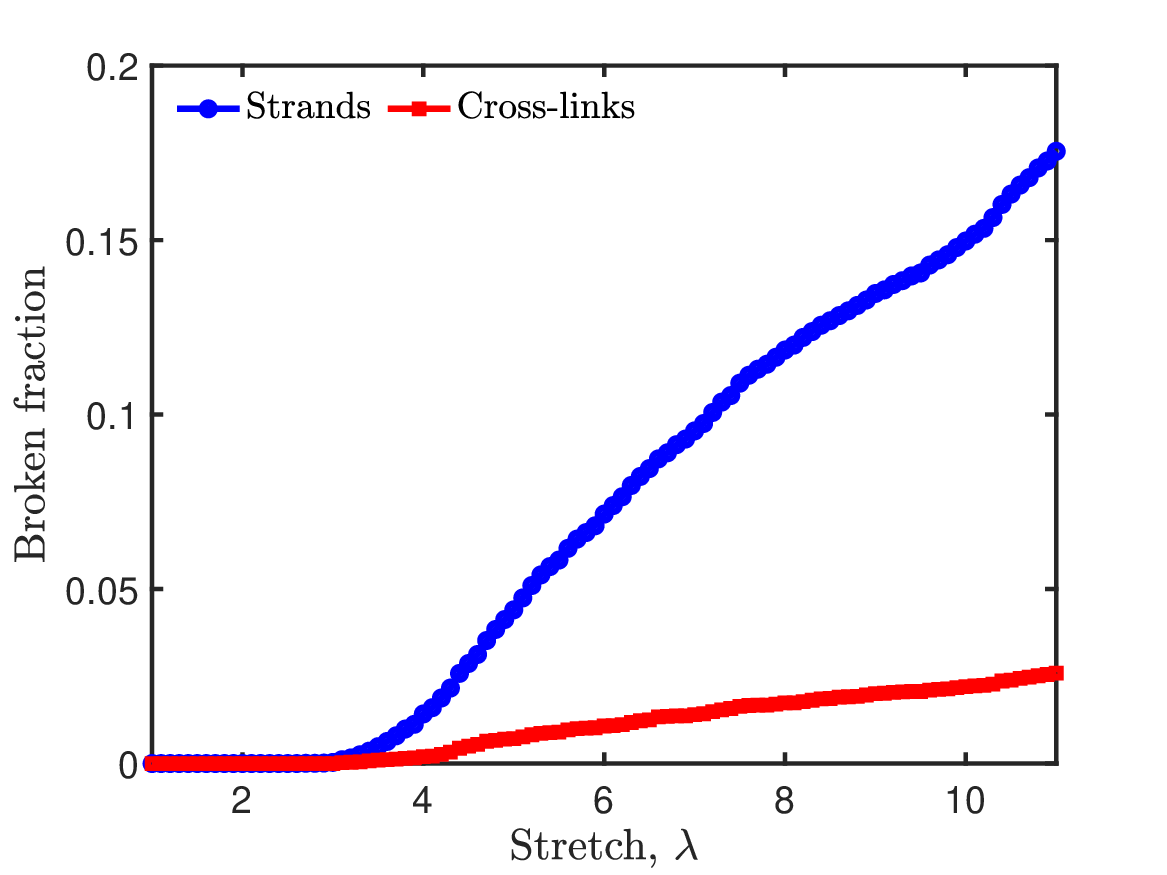}\label{fig:broken_cgmd}}
   \subfigure[]{\includegraphics[width=0.48\textwidth]{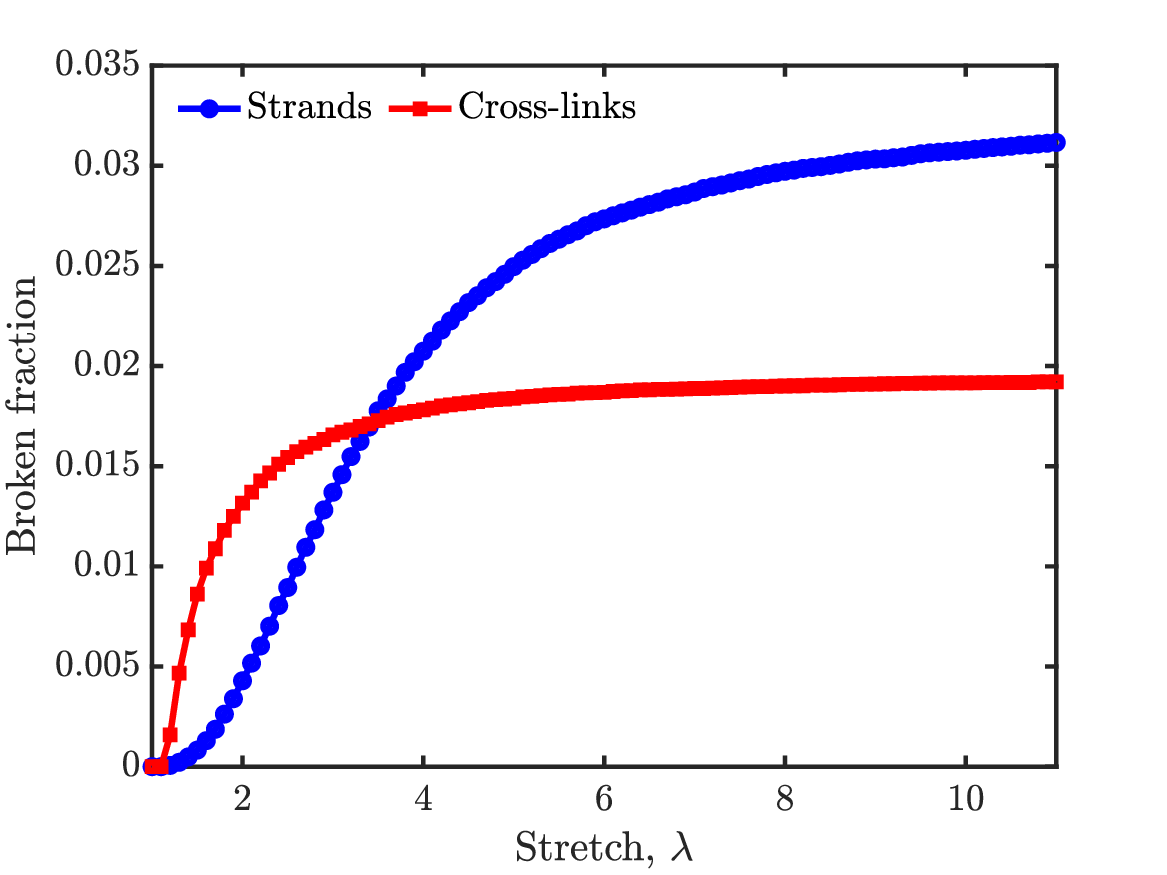}\label{fig:broken_cgsnd}}
    \caption{\textbf{Bond rupture kinetics from CGMD and CGSND.}
Fraction of broken bonds as a function of stretch for polymer strands and cross-links obtained from (a) the CGMD simulations and (b) the CGSND framework.}
    \label{fig:bonds_broken}
\end{figure}
Figure~\ref{fig:bonds_broken} compares bond rupture kinetics obtained from CGMD and CGSND, resolved separately for polymer backbone strands and cross-links. In both approaches, bond rupture initiates only after the network enters the strain-stiffening regime and continues to accumulate throughout post-peak deformation, confirming that fracture is driven by force amplification rather than small-strain elasticity.

Despite this shared qualitative picture, the two models exhibit distinct rupture kinetics. In CGMD (Fig.~\ref{fig:broken_cgmd}), the fraction of broken bonds increases approximately linearly with stretch for both backbone strands and cross-links once rupture begins. Backbone bonds exhibit a consistently higher broken fraction than cross-links, reflecting the larger population of strand bonds available for rupture and the stochastic nature of bond selection in the molecular dynamics simulations.

In contrast, the CGSND framework (Fig.~\ref{fig:broken_cgsnd}) predicts a strongly nonlinear evolution of bond rupture. Both backbone and cross-link rupture fractions rise rapidly at intermediate stretch and then exhibit clear saturation at large deformation. Notably, CGSND predicts an early dominance of cross-link rupture, followed by a crossover to backbone-dominated rupture at higher stretch. This behavior arises from the force-controlled, deterministic rupture criterion in CGSND, which preferentially removes bonds that carry the highest instantaneous forces rather than sampling rupture events uniformly across the bond population.

The differing rupture kinetics highlight a fundamental distinction between the two approaches. In CGMD, bond breaking is mediated by thermal fluctuations and occurs continuously across the available bond population, leading to approximately linear accumulation of damage. In CGSND, rupture is tightly coupled to force localization and load-path evolution, producing a rapid initial increase in rupture probability followed by saturation as the most highly stressed bonds are exhausted and the remaining network carries load more diffusely.

Although the detailed rupture trajectories differ, both approaches exhibit a strong coupling between bond rupture and macroscopic mechanical response. In particular, the rapid increase in rupture rate coincides with the vicinity of the peak stress in both models, indicating that the onset of failure is governed by the emergence of extreme force concentrations rather than by global strain alone. Taken together, these results indicate that CGSND captures the force-driven hierarchy of bond failure, while CGMD reflects the cumulative effects of stochastic rupture across a larger bond ensemble which might weakly capture this hierarchy. These differences motivate a hazard-rate and load localization analysis presented in the following set of results.

\subsection{Rupture kinetics and failure onset}

\begin{figure}[!htbp]
    \centering
    \subfigure[]{\includegraphics[width=0.48\textwidth]{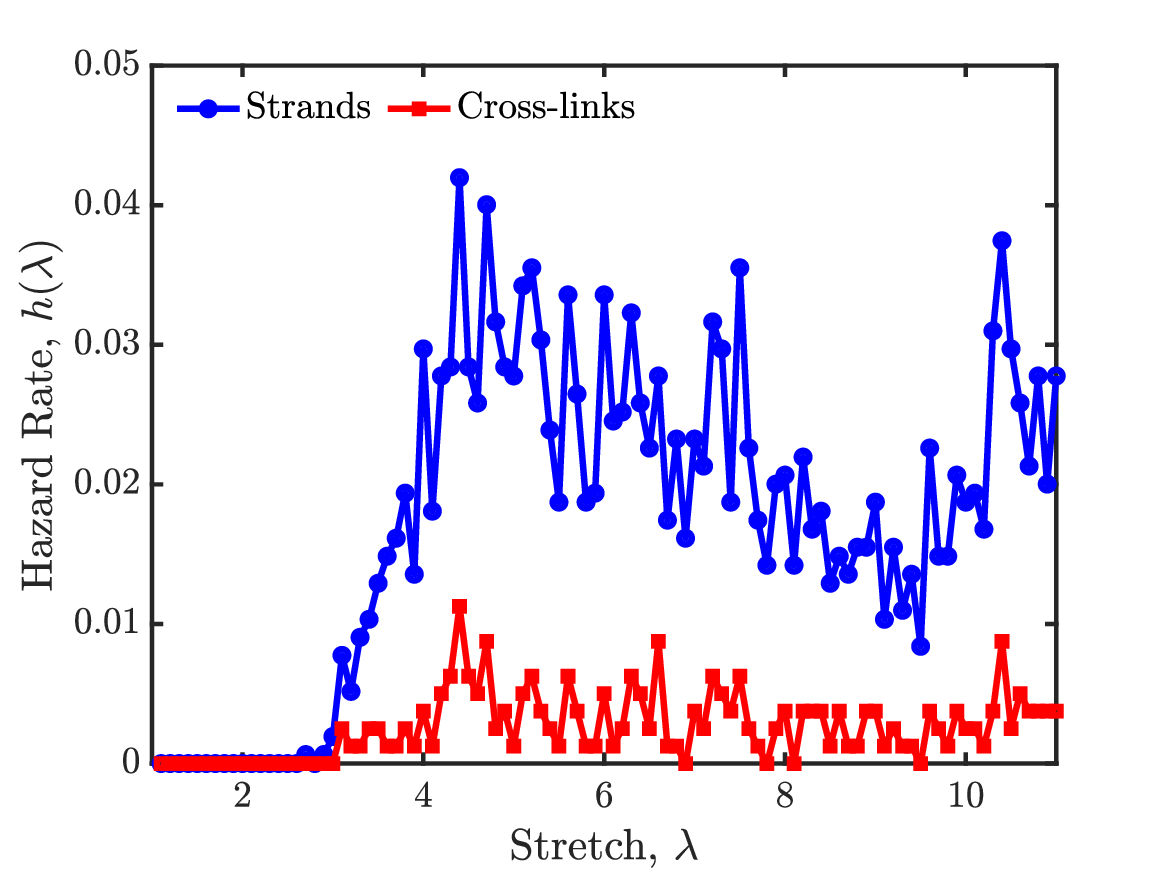}\label{fig:hazard_cgmd}}
   \subfigure[]{\includegraphics[width=0.48\textwidth]{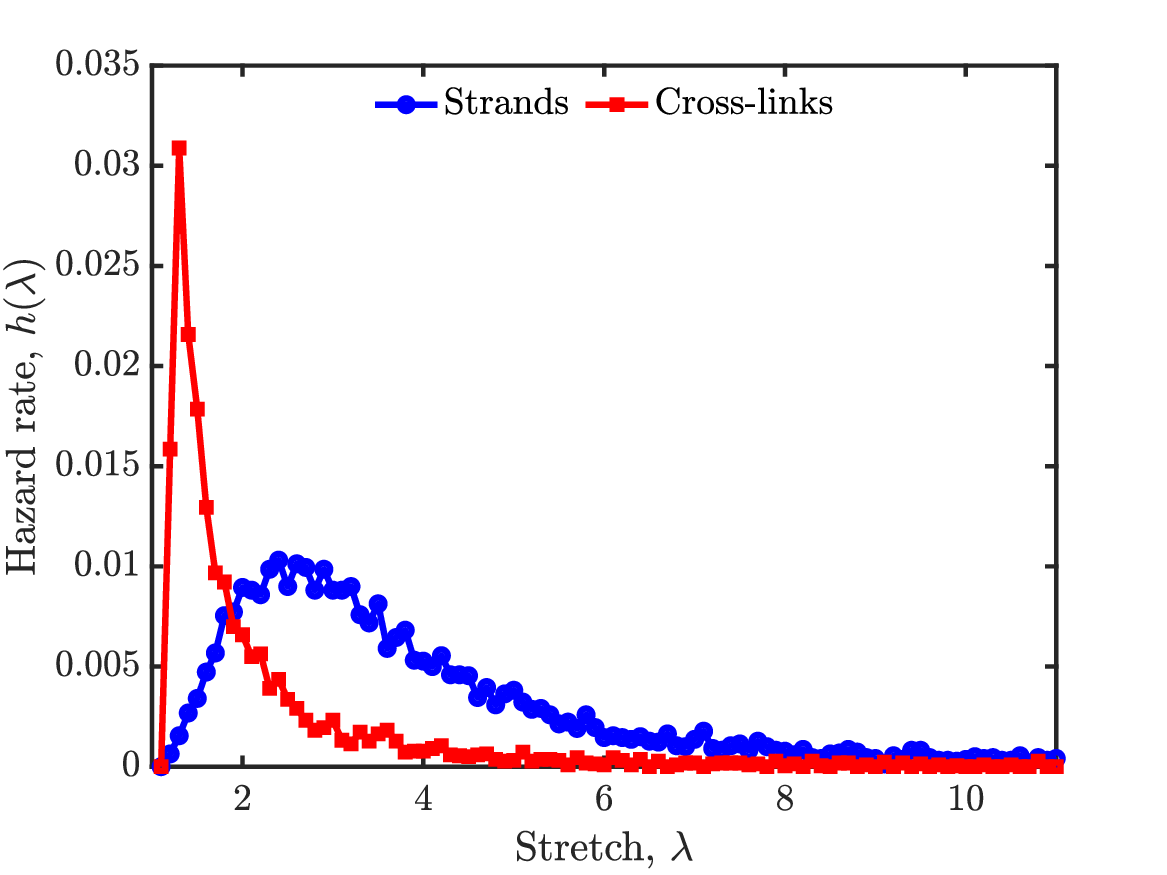}\label{fig:hazard_cgsnd}}
    \caption{\textbf{Rupture kinetics and failure onset.}
Hazard rate of bond-breaking events as a function of stretch for (a) CGMD and (b) CGSND, resolved into cross-link and backbone contributions.}
    \label{fig:hazard_rate}
\end{figure}
Figure~\ref{fig:hazard_rate} reports the rupture hazard rate as a function of stretch for both CGMD and CGSND, resolved into contributions from polymer backbone strands and cross-links. The hazard rate is defined as the instantaneous probability of bond failure per remaining intact bond population and provides a direct measure of the kinetics of damage accumulation.

In both models, the rupture hazard increases with deformation and reaches a pronounced maximum in the vicinity of the ultimate tensile strength. This coincidence indicates that macroscopic stress softening is triggered by a surge in microscopic rupture activity rather than by gradual diffuse damage. The alignment of the hazard-rate peak with the stress maximum provides a kinetic definition of failure onset that is consistent across both modeling frameworks.

Beyond the peak stress, the hazard rate decreases in both approaches, reflecting degradation of the load-bearing backbone and a reduction in the population of strands capable of sustaining high forces. However, the post-peak evolution differs quantitatively between the two models. In CGMD, the hazard rate decays gradually, indicating continued rupture activity driven by thermal fluctuations and dynamical stress redistribution. In contrast, CGSND exhibits a more pronounced post-peak decay, consistent with its force-threshold rupture criterion and affine, rate-independent loading protocol, which rapidly exhausts the subset of strands capable of exceeding the rupture threshold, resulting in accentuated fragmentation of the initial polymer network.

Despite these differences, the presence of a well-defined hazard-rate peak near the ultimate tensile strength in both models demonstrates that failure is governed by a kinetic transition in rupture activity rather than by cumulative bond loss alone. The hazard rate thus provides a physically meaningful and directly comparable measure of failure onset in CGMD and CGSND, bridging microscopic rupture dynamics and macroscopic mechanical response.

\subsection{Distribution of broken bond lengths}

\begin{figure}[!htbp]
    \centering
    \subfigure[]{\includegraphics[width=0.48\textwidth]{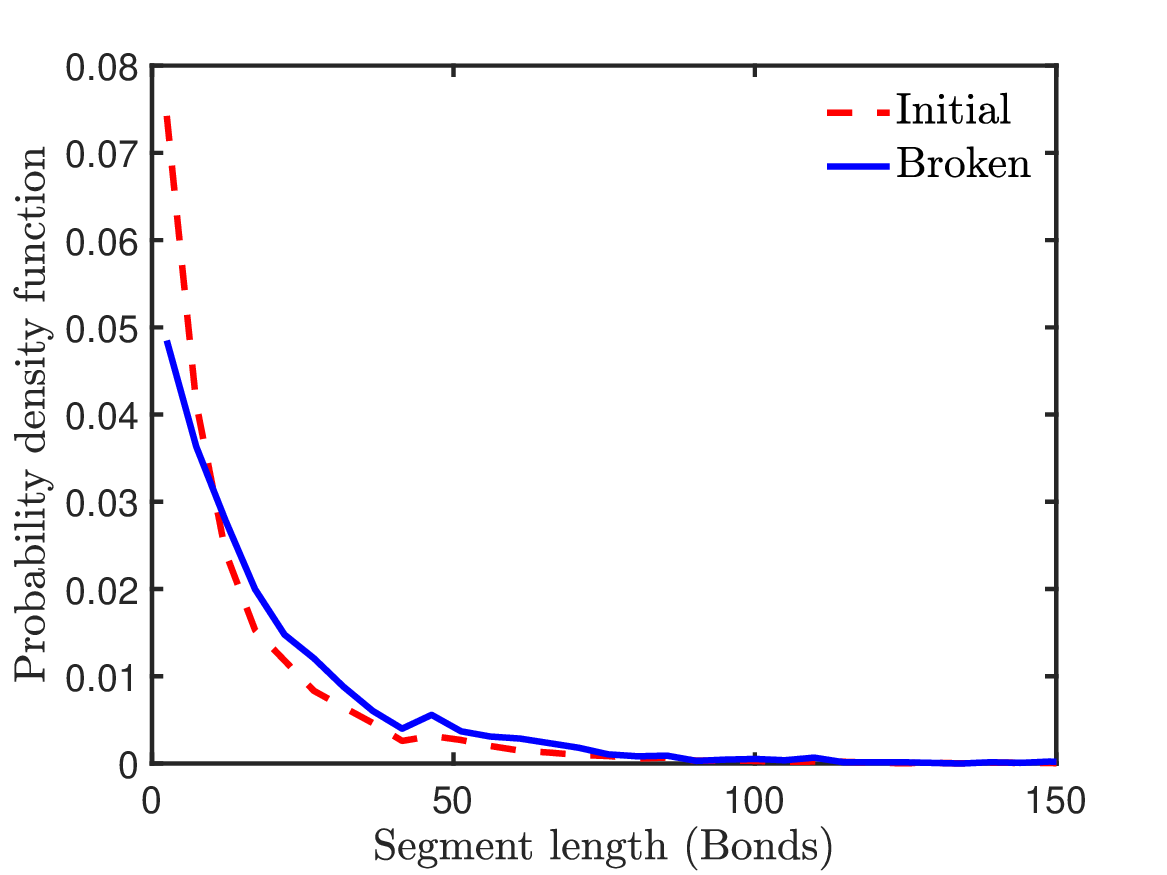}\label{fig:reg_dist_a}}
   \subfigure[]{\includegraphics[width=0.48\textwidth]{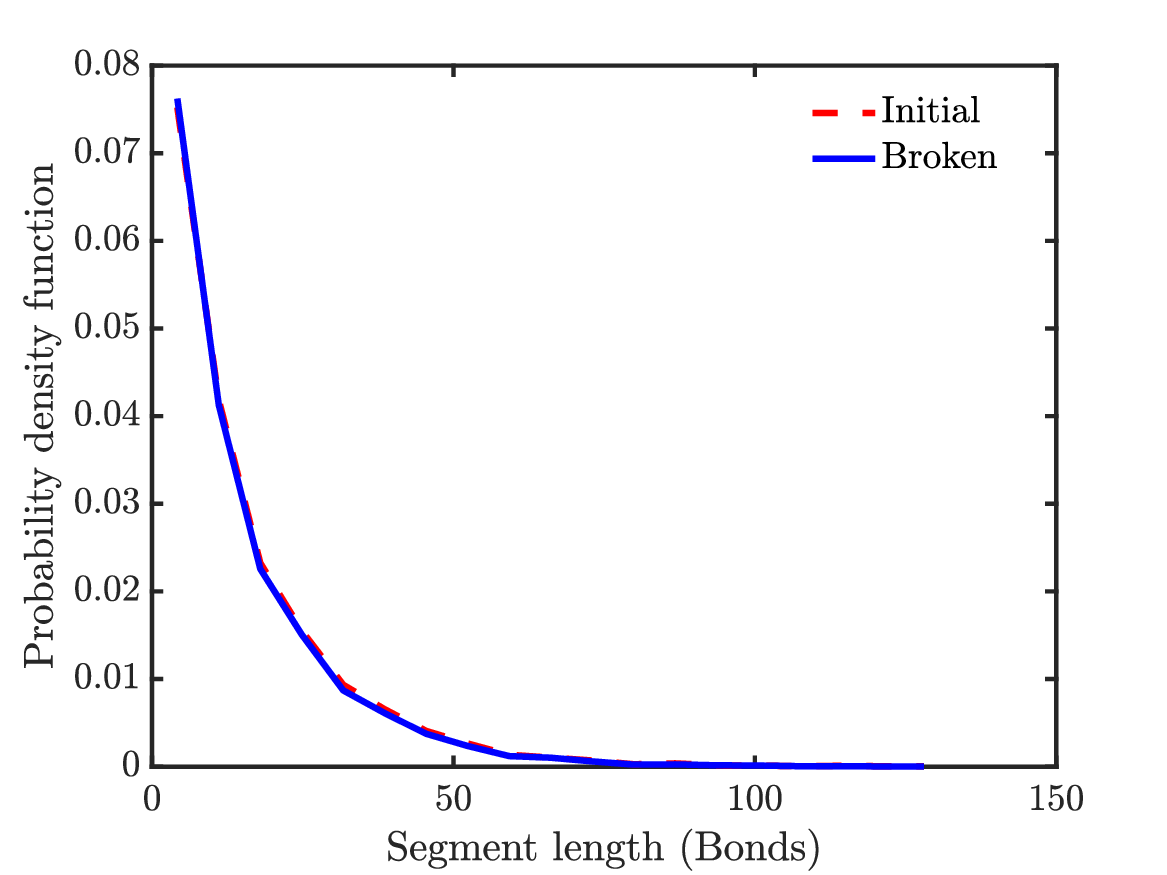}\label{fig:reg_dist_b}}
    \caption{\textbf{Distributions of broken segment contour lengths from CGMD and CGSND.}
Probability density functions of segment contour lengths for the initial network and for segments that rupture during deformation, obtained from (a) CGMD and (b) CGSND.}
    \label{fig:distribution}
\end{figure}
Figure~\ref{fig:distribution} compares the probability density distributions of segment contour lengths among broken bonds with the initial segment length distribution of the undeformed network, for both CGMD and CGSND. In both modeling approaches, the distribution of broken segments closely follows the initial segment population over the entire range of contour lengths.

This result indicates that bond rupture is not preferentially associated with either short or long segments in either CGMD or CGSND. Instead, rupture events sample the underlying segment population broadly, suggesting that failure is governed primarily by network-level stress redistribution and load-path geometry rather than by local segment length alone. In particular, highly stressed load-bearing paths can involve segments spanning a wide range of contour lengths, leading to rupture statistics that reflect the global network structure.

The close correspondence between CGMD and CGSND in the broken-segment length distributions demonstrates that the coarse-grained stochastic network formulation faithfully reproduces the rupture selectivity of the molecular system, despite its simplified, rate-independent dynamics. This observation challenges the commonly held view that shorter chains are intrinsically weaker and more prone to failure, instead supporting a cooperative, network-mediated picture of polymer fracture.

\subsection{Load-path concentration and force localization}

\begin{figure}[!htbp]
    \centering
    \includegraphics[width=0.48\linewidth]{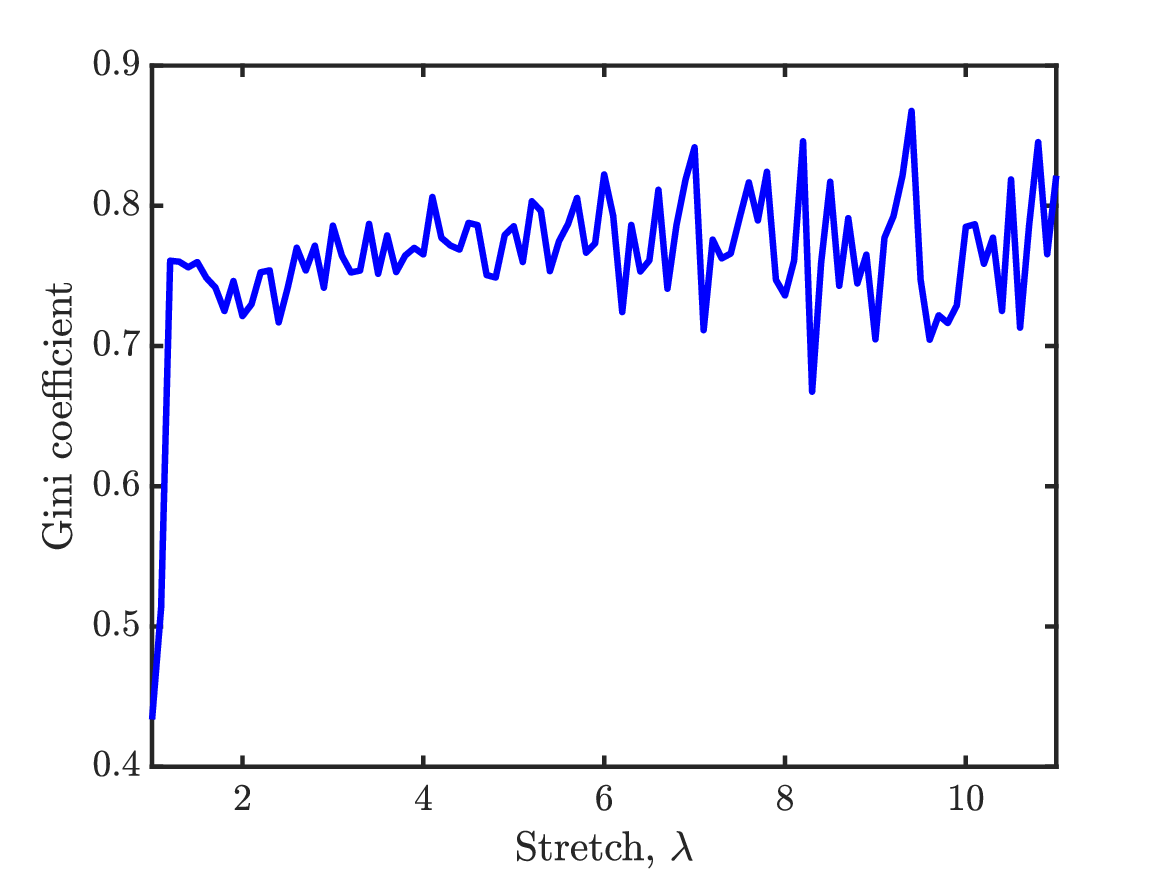}
    \caption{\textbf{Load-path concentration during deformation (CGSND).}
Gini coefficient of bond force magnitudes as a function of applied stretch obtained from the CGSND framework. The Gini coefficient quantifies the degree of force localization within the network, with $G=0$ corresponding to homogeneous load sharing and $G\to1$ indicating extreme localization onto a small subset of strands.}
    \label{fig:gini}
\end{figure}
Figure~\ref{fig:gini} shows the evolution of the Gini coefficient of bond force magnitudes as a function of applied stretch within the CGSND framework. The Gini coefficient provides a scalar, mechanically grounded measure of load-path concentration, directly quantifying the extent to which mechanical load is localized onto a subset of strands.

At small deformation, the Gini coefficient assumes moderate values, reflecting heterogeneous but broadly distributed load sharing across the percolated network. As deformation progresses into the strain-stiffening regime, the Gini coefficient increases rapidly, indicating the emergence of dominant load-bearing paths and strong force heterogeneity. The Gini coefficient reaches a pronounced maximum of approximately $G \approx 0.75$ near the ultimate tensile strength, demonstrating that macroscopic failure is preceded by extreme force localization.

Beyond the peak stress, the Gini coefficient remains largely constant. This constant post-peak phase reflects the insensitivity of force redistribution to progressive rupture within a degrading load-bearing backbone rather than an abrupt collapse of stress transmission. Even after substantial bond breakage, the remaining network continues to route load through a limited subset of surviving strands.

The non-monotonic evolution of the Gini coefficient highlights force localization as a central organizing mechanism underlying failure in polymer networks. Unlike purely topological measures of connectivity, the Gini coefficient captures how mechanical load is distributed and redistributed during deformation and damage accumulation. As a result, it is the force localization, rather than connectivity alone, that governs the onset and progression of macroscopic failure in the CGSND framework.

\section{Discussion} \label{sec:Discp3}

\paragraph{Scope and intent of the model}
The CGSND framework is not intended to reproduce the full dynamical trajectory of polymer networks at the molecular level. Rather, it provides a statistically grounded and mechanically interpretable description of how chain-level entropic elasticity and force-controlled bond rupture combine to generate macroscopic mechanical response. By operating directly on an evolving network representation, the method enables efficient exploration of large system sizes and fracture statistics that are computationally prohibitive for full coarse-grained molecular dynamics (CGMD).
Despite its simplified formulation, CGSND reproduces the key qualitative features of polymer network mechanics observed in CGMD. Most notably, the stress--stretch response exhibits the characteristic nonlinear elastomeric form with a well-defined ultimate tensile strength, followed by post-peak softening driven by progressive bond rupture. The transition from stable load-bearing to rapid failure emerges naturally from the accumulation of force on a diminishing subset of strands.

\paragraph{Failure as a kinetic transition}
Analysis of rupture kinetics reveals that macroscopic failure is governed by a kinetic transition rather than a gradual accumulation of diffuse damage. In both CGMD and CGSND, the bond-breaking hazard rate increases with deformation and exhibits a pronounced peak in the vicinity of the ultimate tensile strength. This peak marks the onset of catastrophic damage accumulation and provides a mechanistically meaningful definition of failure onset that is directly comparable across models.
Beyond the peak stress, the hazard rate decreases as the load-bearing backbone progressively degrades. The sharper post-peak decay observed in CGSND reflects its affine, force-threshold rupture rule, which rapidly exhausts the subset of strands capable of sustaining high forces. In contrast, CGMD exhibits a more gradual decline due to dynamical stress redistribution and continued rupture under finite strain rate conditions.

\paragraph{Rupture selectivity and cooperative failure}
Both CGMD and CGSND show that the distribution of broken segment lengths is statistically indistinguishable from the initial segment population. This observation demonstrates that rupture is not biased toward either short or long chains, contradicting the common assumption that shorter segments are selectively weaker. Instead, failure emerges as a cooperative, network-mediated process governed by global load redistribution and evolving force heterogeneity rather than local segment statistics.

\paragraph{Force localization and load-path concentration}
The evolution of the Gini coefficient of bond force magnitudes provides a complementary, mechanically grounded perspective on failure. In CGSND, the Gini coefficient rises sharply during strain stiffening, peaks near the ultimate tensile strength, and decays slowly in the post-peak regime. This non-monotonic behavior indicates that extreme force localization precedes macroscopic failure and persists even as the network degrades.
Unlike purely topological measures, the Gini coefficient directly quantifies how stress is routed through the network and how this routing evolves during damage accumulation. The persistence of high Gini values beyond the stress maximum demonstrates that failure proceeds through progressive exhaustion of dominant load paths rather than an abrupt loss of connectivity.

\paragraph{Advantages and limitations}
Taken together, these results demonstrate that CGSND provides both computational efficiency and physical interpretability. Its principal advantages are: (i) the ability to capture nonlinear stress--strain response and failure onset without resolving microscopic dynamics, (ii) direct access to rupture kinetics and force localization through hazard rates and Gini coefficients, and (iii) transparent mechanistic links between microscopic bond failure and macroscopic load-bearing capacity.
At the same time, important limitations remain. The method does not resolve local segmental relaxation, inertial effects, or thermal fluctuations, and therefore represents an affine, rate-independent limit of the underlying molecular system. In contrast, CGMD corresponds to a finite strain-rate process in which nonaffine rearrangements and thermostat-mediated dissipation contribute to stress redistribution. Quantitative differences in stress magnitude are therefore expected even when the qualitative failure mechanisms are approximately reproduced.

\paragraph{Outlook}
Despite these limitations, the CGSND framework provides a promising route for bridging network-level statistical physics and coarse-grained mechanical modeling. By combining entropic elasticity, force-controlled rupture, rupture kinetics, and force localization within a unified network formulation, the method enables computationally tractable studies of fracture, heterogeneity, and scaling in polymer networks across system sizes inaccessible to molecular simulation alone.
\section{Conclusion} \label{sec:Concp3}

We have introduced a coarse-grained stochastic network dynamics (CGSND) framework for modeling deformation and rupture in polymer networks. Despite its simplified, rate-independent formulation, the method reproduces the key qualitative hallmarks of elastomeric mechanics: a nonlinear stress--stretch response with a well-defined ultimate tensile strength, a sharp transition in rupture kinetics, and progressive post-peak softening driven by bond failure.
Direct comparison with coarse-grained molecular dynamics demonstrates that macroscopic failure is governed by a kinetic transition in bond-breaking activity rather than gradual diffuse damage. The emergence of a pronounced peak in the rupture hazard rate near the ultimate tensile strength provides a mechanistically meaningful marker of failure onset. In addition, the statistical equivalence of initial and broken segment length distributions shows that rupture is not selectively biased toward short chains, but instead arises as a cooperative, network-mediated process.

By operating directly on an evolving graph representation, CGSND establishes transparent links between chain-level force accumulation, rupture kinetics, and macroscopic mechanical response, while remaining computationally efficient enough for large-scale statistical studies. At the same time, the absence of explicit segmental relaxation and thermal dynamics places the framework in an affine, rate-independent limit relative to molecular simulation, clarifying the origins of quantitative differences in stress magnitude.
Looking forward, extensions incorporating bond-energy heterogeneity, stochastic activation, or healing mechanisms would enable systematic studies of fracture, fatigue, and reliability in polymer networks across length and time scales. More broadly, this work demonstrates how network-based coarse-graining can bridge statistical physics descriptions of disordered systems with continuum mechanical behavior, providing a scalable and physically interpretable framework for studying failure in complex polymeric materials.

\section*{Data Availability} 
The source code and the analysis can be made available upon request to the authors.
\section*{Conflict of Interest} 
The authors declare no competing interests.
\section*{Acknowledgments} 
S.M. and W.C. acknowledge support from the Air Force Office of Scientific Research under award number FA9550-20-1-0397.

\bibliographystyle{unsrtnat} 
\bibliography{ref_p3}
\end{document}